\documentclass[onecolumn,showpacs,preprintnumbers,amsmath,amssymb]{revtex4}

\usepackage{graphicx}
\usepackage{dcolumn}
\usepackage{bm}
\usepackage{textcomp}
\usepackage{wasysym}
\usepackage{stmaryrd}
\usepackage{subfig}
\usepackage{color}

\def \ee{\end{equation}}
\def \be{\begin{equation}}
\def \bea{\begin{eqnarray}}
\def \eea{\end{eqnarray}}

\preprint{}
\begin{document}

\title{Closing a window for massive photons}

\keywords      {}
\author{Sergio A. Hojman* and Benjamin Koch**}
 \affiliation{
*Departamento de Ciencias,
Facultad de Artes Liberales, Facultad de Ingenier\'{\i}a y Ciencias,
Universidad Adolfo Ib\'a\~nez, Santiago, Chile,\\ and Departamento
de F\'{\i}sica, Facultad de Ciencias, Universidad de Chile,
Santiago, Chile,\\ and Centro de Recursos Educativos Avanzados,
CREA, Santiago, Chile; \\
 ** Pontificia Universidad Cat\'{o}lica de Chile, \\
Av. Vicu\~{n}a Mackenna 4860, \\
Santiago, Chile; \\
}
\date{\today}

\begin{abstract}
Working with the assumption of non-zero photon
mass and a trajectory that is 
described by the non geodesic world line of a spinning top
we find, by deriving new astrophysical bounds,
that this assumption is in contradiction with
current experimental results. 
This yields the conclusion that such photons have to
be exactly massless. 
\end{abstract}

\pacs{04.62.+v, 03.65.Ta}
\maketitle



%
\section{Introduction}
Although there are good theoretical reasons to
believe that the photon mass should be exactly zero,
there is no experimental proof of this belief.
A long series of very different experiments
lead to the current experimental upper bound
on the photon mass $m_\gamma<10^{-18} eV$.
However, even with further improvement of the
experimental technology and precision a complete
exclusion of a non-zero photon mass by those techniques
will never be possible.
In order to improve on this situation we will work
with the assumption that the photon mass is different
from zero $m_\gamma \neq 0$. 
We will not speculate on the origin of this mass and
its underlying theory. In the simplest case, such a photon
will be described by a wave equation of the Proca type.
For waves describing massless particles with spin, it is well understood,
in the eikonal approximation,
how the propagating wave can be treated 
as geometric path that minimizes a Lagrangian described by the length
\be\label{lag1}
g_{\mu \nu}u^{\mu}u^{\nu}=u^2\quad,
\ee
where $u^{\mu}=\dot x^\mu$. 
As soon as mass and rotational degrees of freedom ($m^2$, $\sigma^{\mu\nu}$) are involved,
it is natural to assume, that in the eikonal approximation, the resulting geometric action
also involves the invariant functions of the corresponding additional degrees of freedom
\bea\label{lag2}
g_{\mu \nu}g_{\alpha \beta}\sigma^{\alpha\mu}\sigma^{\nu \beta}
&=&tr \sigma^2\quad, \\ \nonumber
g_{\mu \nu}g_{\alpha \beta}g_{\gamma\delta}u^\mu \sigma^{\nu \alpha}
\sigma^{\beta \gamma}u^\delta
&=&u\sigma\sigma u\quad, \\ \nonumber
g_{\mu \nu}g_{\rho \tau}g_{\alpha \beta}g_{\gamma\delta}
\sigma^{\delta \mu}\sigma^{\nu \rho}\sigma^{\tau \alpha}\sigma^{\beta \gamma}&=&
tr\sigma^4\quad.
\eea
This generic approach to the geometric action 
which includes spin and mass was
developed and discussed in the context of a spinning top, a point on a world line
to which a rotating frame has been attached
\cite{mat,Papapetrou:1951pa,Corinaldesi:1951pb,hr,Hojman:1978yz,hojman1,mas,Hojman:1978wk,Hojman:2012me}.
The resulting equations of motion do not depend on the particular form in which
the Lagrangian terms (\ref{lag1},\ref{lag2})  are combined.
Exact solutions to those equations coupled to gravity via the metric $g_{\mu \nu}$ 
show that spinning tops without mass follow geodesics, while massive spinning tops
have different trajectories.
Further implications and discussions of the spinning top approach 
can be found in
\cite{Wald:1972sz,Barker:1975ae,Barducci:1976xq,Hehl:1976kj,Piriz:1996mu,Bailey:2010af}.

\section{Astrophysical bound}
As master equation for the trajectory 
of massive particles with spin we use the solution of the spinning top
equations in a Scharzschild background
\cite{hr,Hojman:1978yz,Hojman:1978wk,Hojman:2012me}
\begin{equation}
\frac{d\phi}{dr}=\left(\frac{2\eta+1}{\eta-1}\right)\left(\frac{P_\phi}{r^2
P^r}\right)\, .
\label{dphidr}
\end{equation}
where one has to insert the following definitions
\begin{equation}
 \eta=\frac{J^2 r_0}{2 m_\gamma^2c^2 r^3}\, ,
\label{eta}
\end{equation}
where $J=\hbar$ is the photon spin. For a given
angular momentum $j$, the momenta are
\begin{equation}
 P_\phi=\frac{-j+(\pm_\phi) E J/(m_\gamma c^2)}{1-\eta}\, ,
\label{pfi}\end{equation}
\begin{equation}
 P_t=\frac{E -(\pm_t) j J r_0/(2 m_\gamma r^3)}{1-\eta}\, ,
\label{pt}\end{equation}
and 
\begin{equation}
 P^r=\pm
\left[\frac{P_t^2}{c^2}-\left(\frac{P_\phi^2}{r^2}+m^2c^2\right)
\left(1-\frac{r_0}{r}\right)\right]^{1/2}\,.
\label{pr1}\end{equation}
The orientation of this problem is chosen such
that the angle  $\theta=\pi/2$ is constant
along the trajectories, which implies
\be
P_{\theta}=0\quad.
\ee
The solution is such that $\pm_\phi=\pm_t$
and $P_\mu P^\mu=m_\gamma^2c^2$ are fulfilled.

For small spin corrections and large radii one can approximate
\be\label{drdphiapprox}
\left(\frac{dr}{d\phi}\right)^2=r^4
\left(\frac{1}{j^2}+(\pm_\phi) \frac{2\hbar}{c j^3 m_\gamma}\right)-r^2+r r_0 
\left(1-\frac{(\pm_t) \hbar }{c j\, m_\gamma}\right)\quad.
\ee
In this equation one can easily verify that for
spinless case ($\hbar\rightarrow 0$),
the geodesic trajectory is recovered.
One can express the angular momentum $j$ in terms
of the minimal radial distance $r_m$ 
\begin{small}
\bea\label{jm}
j &=& \frac{2 c E \hbar m \pm (3 r_0-2 r_{m}) r_{m}^4}
{c \hbar^2 r_0^2 r_{m}+4 c^3 m^2 (r_0-r_{m}) r_{m}^4}+
2 \sqrt{r_m(r_m-r_0)}\cdot \\ \nonumber &&
\frac{\sqrt{
  E^2 \hbar^4 r_0^2 r_{m}^4+4 c^8 m^6 (r_0-r_{m}) r_{m}^9+
c^6 \hbar^2 m^4 r_0 r_{m}^6 \left(4 r_{m}-3 r_0\right)+c^2 \hbar^2
 r_0 \left(\hbar^4 r_0^3/4-4 E^2 m^2 r_{m}^7\right)+ c^4 m^2 r_{m}^3 \left(4 E^2
m^2 r_{m}^7-\hbar^4 r_0^2 r_{m}\right)}}
{c \hbar^2 r_0^2 r_{m}+4 c^3 m^2 (r_0-r_{m}) r_{m}^4}
\eea
\end{small}
Due to the steep behavior of $d\phi/dr$ at the minimal radius $r_m$
one has to use (\ref{jm}), since one
cannot rely a priory 
on the approximate value $j\approx r_m^{3/2}/\sqrt{r_m-r_0}$.

\begin{figure}[hbt]
   \centering
\includegraphics[width=12cm]{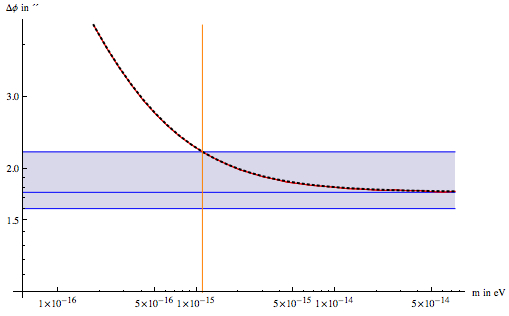}
  \caption{\label{figdelfi}
 Numerical results for the angular deviation $\Delta \phi$ as a function
 of $m_\gamma$.
 The horizontal lines represent a conservative range for the
 measured gravitational lensing by the sun \cite{Weinberg}. Please note
 that this range can be largely improved by measurements using radio astronomy 
 \cite{Will,Fomalont:2009}.
The red line is the numerical
 value for both configurations of $\pm_\phi=\pm_t$. The black
 dotted line is the analytic estimate (\ref{devi}). 
  The vertical line is the deduced minimal
 value for $m_\gamma$ given in (\ref{exp1}). }
\end{figure}

Due to the involved form of (\ref{jm}), the 
complete angular deviation has to be computed
essentially numerically, as it was done here. 
In order to gain a better intuition of the numerical results 
one can however do approximations
and expansions for small $\hbar$, $m c^2/E$, and $r_0/r_m$
\be\label{devi}
\Delta \phi=2 (\int_{r_m}^\infty \left(\frac{d\phi}{dr}\right) dr)-\pi
\approx2 \frac{r_0}{r_m}
\left(1+\frac{\hbar }{r_m c \,m_\gamma} \right)\quad,
\ee
where $\pm_t=-1$ and $\pm_\phi=1$ was used.
One observes that the standard result $2 r_0/r_m$ is modified by a spin-dependent
correction. 
In order to be in agreement with the experimentally well confirmed
gravitational lensing effect, this dimensionless modification has to be much smaller than one
\be\label{DeltaJ}
1\gg\Delta_J\approx \frac{\hbar }{r_m c \,m_\gamma}\quad.
\ee
This modification is inversely proportional to the
photon mass $m_\gamma$. 
This inverse proportionality implies that practically 
no deviations from the usual trajectories are
observable for massive standard model particles.
Please note that this non perturbative
feature in $m_\gamma$ is also known from theories of massive gravity where
the limit $m_g \rightarrow 0$ does not give standard gravity
where the graviton was massless right from the start $m_g=0$.
Further similarities to massive gravity have been recently discussed in \cite{Deser:2012qx}.
It is further interesting to
note that $\Delta_J$ actually increases the usual angular deflection. 
However, a deviation from the 
geodesic bending of light has been excluded to high precision \cite{Fomalont:2009}.
Thus, the relation (\ref{DeltaJ}) can be interpreted as an estimate for 
the numerical lower photon mass limit
\be
m_\gamma\gg \frac{\hbar }{r_m c }\approx3\times 10^{-16}\,eV/c^2 \quad.
\ee
In figure \ref{figdelfi} we compare this estimate
to the observed value \cite{Weinberg,Will}
and to the precise numerical results by using
the solar radius $r_m=6.96\times 10^8$~m, the solar
Schwarzschild radius $r_0=2964$~m, and a photon energy of $E=1$~eV.
One finds that the estimated deviation is in very good agreement
with the lower numerical result. 
By using conservative exclusion ranges for $\Delta \phi$
one can read from figure a more precise limit 
\be\label{exp1}
m_\gamma > 1.1\times 10^{-15}\,eV/c^2\quad.
\ee 
%
The lower limit (\ref{exp1}) can be combined with the upper limit for a photon mass  
from the Particle Data Group \cite{PDG}
$m_\gamma<10^{-18}\,eV/c^2$.
Some of the limits in the PDG tables \cite{PDG} are however derived
by assuming a spinless coupling of the photon to gravity and
to matter. But this might not be true. When deriving the limit (\ref{exp1}),
we found the gravitational coupling of the massive particles with spin,
is different from a spinless or mass-less coupling.
Thus, out of the limits in \cite{PDG} only those can be applied
straightforwardly, where the gravitational coupling is not relevant.
For example, results from laboratory experiments, that do not involve
astrophysical or gravitational components 
\cite{Spavieri:2011zz,Franken:1971mq,Williams:1971ms,Chernikov:1992sb,Accioly:2010zzc} 
can be used directly. Also results from other experiments,
that do involve astrophysical components but where the actual trajectory
of the photon does not play any role 
\cite{Lou:2003,Ryutov:2009zz,Lakes:1998mi,Davis:1975mn,Hollweg:1974mp,Fischbach:1994ir,Gintsburg:1964my,Patel:1965mu,Goldhaber:2008xy}
should be applicable, but a sound revision is in order.
Nevertheless, the bounds on $m_\gamma$ from some experiments are not
directly applicable 
because they are not sufficiently general \cite{Chibisov:1976mm,Adelberger:2003qx},
or because they make explicit use of the photon trajectory in a gravitational field
\cite{Accioly:2010zzc,Fullekrug:2004zz,Rob91,Shapiro:2004zz,Lambert:2009xy}
(Note that those experimental references are ordered by the strength of their
respective bounds on $m_\gamma$).
The experimental bounds, that are directly applicable are\\
$m_\gamma<$
$\{1.2\times 10^{-18}\,eV/c^2$ \cite{Spavieri:2011zz},
$5.6\times 10^{-17}\,eV/c^2$ \cite{Franken:1971mq},
$1\times 10^{-14}\,eV/c^2$ \cite{Williams:1971ms},
$4.5\times 10^{-10}\,eV/c^2$ \cite{Chernikov:1992sb},
$1.6\times 10^{-4}\,eV/c^2$ \cite{Accioly:2010zzc}$\}$.

Thus, three of the applicable 
experimental limits \cite{Williams:1971ms,Chernikov:1992sb,Accioly:2010zzc},
combined with the limit (\ref{exp1}), leave only a window for the
photon mass.
However, if one refers only to the more stringent upper limits on $m_\gamma$  
\cite{Spavieri:2011zz,Franken:1971mq} one 
finds
\be\label{exp2}
m_\gamma < \{ 1.2\times 10^{-18}\,eV/c^2,\; 5.6\times10^{-17}\,eV/c^2 \} \quad,
\ee
which is in clear contradiction to (\ref{exp1}).
Therefore, one can conclude that confronting the assumption of a non-zero
photon mass with an astrophysical lower bound and the established
upper bounds excludes the existence of any photon mass.
Thus, the photon has to be massless right from the start
\be\label{concl}
m_\gamma=0\quad.
\ee
At this point, a word of caution is at place:
When combining bounds that arise from different descriptions,
as it was done here when combining the bounds from a geometrical description (\ref{exp1})
with bounds that arise from a wave description (\ref{exp2}), one
might end up comparing parameters of different models. Even
within the upper bounds obtained from wave descriptions of the massive
photons exist problems of generality, which imply that bounds that were
observed for one model of $m_\gamma$ do not apply to other models of
$m_\gamma$ \cite{PDG}. 

Thus, despite of the generality of the geometric equations of motion,
the criterion of applicability that was
used here, can be made more precise by explicitly deriving specific geometrical
equations as eikonal limit of a particular wave model of photon mass \cite{SHBK}.
A useful guiding tool for anticipating those results might be the peculiar non continuous 
behavior in the $m_\gamma \rightarrow 0$ limit of the geometric description,
as it appears in equation (\ref{DeltaJ}).
For example for the wave descriptions of $m_\gamma$
that use the Proca equations \cite{Proca:1936},
it has been shown that the limit $m_\gamma \rightarrow 0$
converges to Maxwell's theory only if one demands 
$\partial_\mu A^\mu=0$ 
\cite{Schrod:1955,Coester:1951,Ume:1953,Glauber:1953,Stueckelberg:1957,Boulware:1962}.
On the other hand, the same limit for Proca equations
with $\partial_\mu A^\mu\neq 0$ does not lead continuously to Maxwell's 
 theory \cite{Deser:1972,Boulware:1989up}.
The corresponding geometrical descriptions
in the eikonal limit are expected to have an analogous behavior in the 
$m_\gamma \rightarrow 0$ limit as their counterpart in the wave description.
However, the eikonal limit of Maxwells theory is described by the usual geodesics.
Therefore, one might expect that the limit (\ref{exp1})
could only apply to Proca mass models with $\partial_\mu A^\mu\neq 0$
and not to Proca mass models with $\partial_\mu A^\mu= 0$.
However, this argument does not disqualify the
applicability of the geometric approximation (\ref{lag2}) a priory.
For example, in \cite{Deser:1972} it has been shown that, although the 
longitudinal modes of the Proca equation
(even with $\partial_\mu A^\mu= 0$) decouple from matter
for $m_\gamma \rightarrow 0$ in flat space-times, the coupling
to gravity of those modes does not vanish 
in this limit.

\section{Discussion and Conclusion}
Under the assumption that the trajectory of a photon 
can be described by the world line of a top with spin one, 
it was shown that such a photon has to have either zero mass $m_\gamma=0$
or a minimal non-zero mass $m_\gamma >1.1\times 10^{-15}\,eV/c^2$, in order to be in agreement 
with the observed gravitational light bending by the sun 
\cite{Weinberg,Will,Fomalont:2009}. Combining those results
with the experimental bounds from laboratory experiments such as 
 $m_\gamma<5.6\times10^{-17}\,eV/c^2$ \cite{Spavieri:2011zz,Franken:1971mq},
one finds that, in this theoretical framework, there is no room for a photon mass different from zero.
 \\

The work of B.K. was supported proj. Fondecyt 1120360 and Anillo Atlas Andino 10201.
Many thanks to S.~Deser for his helpful remarks.



\end{document}